\documentclass{aa}  
\usepackage{graphicx}
\usepackage{natbib}
\usepackage{txfonts}
\usepackage{longtable}
\usepackage{supertabular}
\begin{document}
   \title{The HARPS search for southern extra-solar planets}

   \subtitle{IX. $\mu$\,Ara, a system with four planets\footnote{Based on observations made with the HARPS instrument on ESO's 3.6\,m telescope at the La Silla Observatory in the frame of the HARPS GTO programme ID $<$072.C-0488$>$ and under programme ID $<$073.D-0578$>$}  }

   \author{F.~Pepe\inst{1}
          \and
           A.C.M.~Correia\inst{2} 
	  \and
	   M.~Mayor\inst{1}
	  \and
          O.~Tamuz\inst{1}
	  \and
          W.~Benz\inst{3}
	  \and
	  J.-L.~Bertaux\inst{4}
	  \and
          F.~Bouchy\inst{5}
	  \and
	  J.~Couetdic\inst{6}
	  \and
	  J.~Laskar\inst{6} 
	  \and
          C.~Lovis\inst{1}
	  \and
           D.~Naef\inst{7,1}
	  \and
           D.~Queloz\inst{1}
	  \and
            N.~C.~Santos\inst{8,1} 
	  \and
          J.-P.~Sivan\inst{5}
	  \and
          D.~Sosnowska\inst{1}
	  \and
          S.~Udry\inst{1}	  	  
          }

   \offprints{\newline F. Pepe, \email{Francesco.Pepe@obs.unige.ch}}

   \institute{
          Observatoire de Gen\`eve, 51 ch. des Maillettes, CH--1290 Sauverny, Switzerland
          \and
	  Departamento de F\'\i sica da Universidade de Aveiro, Campus
	  Universit\'ario de Santiago, P--3810-193 Aveiro, Portugal
	  \and
	  Physikalisches Institut Universit\"at Bern, Sidlerstrasse 5, CH--3012 Bern, Switzerland
	  \and
	  Service d'A\'eronomie, Route des Gatines, F--91371 Verri¶res le Buisson CÚdex, France
	  \and
          Laboratoire d'Astrophysique de Marseille, Traverse du Siphon BP8, F--13376 Marseille CEDEX
           12, France
	  \and
	  Astronomie et Syst\`emes Dynamiques, IMCCE-CNRS UMR 8028, 77 Avenue
	  Denfert-Rochereau, F--75014 Paris, France
	  \and
	  ESO La Silla Observatory, Alonso de Cordova 3107, Vitacura Casilla 19001, CL--Santiago 19,
	   Chile
	  \and
	  Observat\'orio Astron\'omico de Lisboa, Tapada da Ajuda, P--1349-018 Lisboa, Portugal
	     }

   \date{received; accepted}

    \abstract
   {}
   {The $\mu$\,Ara planetary system is rather complex: It contains two already known planets, $\mu$\,Ara\,b with $P=640$\,days and $\mu$\,Ara\,c with $P=9.64$\,days , and a third companion on a wide but still poorly defined orbit. Even with three planets in the system, the data points keep anomalously high dispersion around the fitted solution. The high residuals are only partially due to the strong p-mode oscillations of the host star. We have therefore studied in this paper the possible presence of a fourth planet in the system.}
   {During the past years we have carried out additional and extremely precise radial-velocity measurements with the HARPS spectrograph. These data turned out to be of high importance for constraining the many free parameters in a four-planet orbital fit. Nevertheless, the search for the best solution remains difficult in this complex and multi-dimensional parameters space. The use of the new \emph{Stakanof} software and employing an optimized genetic algorithm, helped us considerably in this task and made our search extremely efficient and successful.}
   {We provide in this paper a full orbital solution of the planetary system around $\mu$\,Ara. It turns out to be the second system known to harbor 4 planetary companions. $\mu$\,Ara\,b was already well known and characterized before this study. Thanks to the new data points acquired with HARPS we can confirm the presence of $\mu$\,Ara\,c at $P=9.64$\,days, which produces a coherent RV signal over more than two years. The new orbital fit sets the mass of $\mu$\,Ara\,c to 10.5\,M$_{\oplus}$. Furthermore, we present the discovery of $\mu$\,Ara\,d, a new planet on an almost circular 310\,days-period and with a mass of 0.52\,M$_{Jup}$. Finally, we give completely new orbital parameters for the longest-period planet, $\mu$\,Ara\,e. It is the first time that this companion is constrained by radial-velocity data into a dynamical stable orbit, which leaves no doubt about its planetary nature. We take this opportunity to discuss naming conventions for poorly characterized planets.   
    } 
   {}


 \maketitle

%
\section{Introduction}

The discovery of the first planet around  $\mu$\,Ara (HD\,160691) has been reported by \citet{Butler:2001} in 2001. At that time, the authors announced a 1.97\,M$_{Jup}$-mass planet orbiting its host star in $P=743$\,days on a very eccentric orbit ($e=0.62$). The acquisition of new data points during the following year called however for a significant correction of the preliminary parameters of $\mu$\,Ara\,b: \citet{Jones:2002} indicated a new orbital solution with $P=638$\,days and much lower eccentricity $e=0.31$.  From a clear trend in the radial-velocity data the same authors also deduced the presence of a second companion, for which the orbital parameters were however unconstrained. Their data set only a lower limit for the period and the minimum mass, leaving the possibility for a stellar companion still open. The $\mu$\,Ara system clearly deserved further investigations.

\par
Indeed, asteroseismology observations, which aimed at testing whether its high metallicity was of primordial nature or rather the product of planet engulfment \citep{Bouchy:2005}, unveiled the presence of $\mu$\,Ara\,c, a very low-mass planet of only 14\,M$_{\oplus }$ on a 9\,days orbit \citep{Santos:2004-b}. The low radial-velocity wobble amplitude of 4\,m\,s$^{-1}$ was detected thanks to the extreme precision of the HARPS instrument \citep{Mayor:2003} and to the high measurement density. In fact, the stellar pulsation modes, which, on $\mu$\,Ara may attain amplitudes up to 6 to 9\,m\,s$^{-1}$ p-p, must be averaged out by integrating over times scales considerably longer than the pulsation period. Under these conditions, the HARPS asteroseismology run provided data points with night-to-night dispersion below 0.5\,m\,s$^{-1}$. The Neptune-mass planet $\mu$\,Ara\,c was also confirmed by a continuous follow-up with the HARPS instrument during the following months.

\par
At about the same time \citet{McCarthy:2004} published a possible orbital solution for the long-period companion, although, in their analysis, they did not take into account the presence of the innermost planet. The single Keplerian with a linear trend was no longer fitting the new data points correctly. They showed that the residuals were significantly reduced when assuming instead a two-Keplerian model with the second companion of 3.1\,M$_{Jup}$ orbiting the parent star in 8.2\,years on a eccentric orbit with $e=0.57$. Since the long-period planet orbit was only covered partially, large uncertainties still persisted on the orbital parameters of the outer planets. A recent dynamical analysis of the $\mu$\,Ara system performed by \citet{Gozdziewski:2004} showed that the published orbit was dynamically unstable and confirmed that the systems was still unconstrained, leaving room towards longer periods and more massive companions. By investigating possible stable orbital solutions within a 3$\sigma$ range from the best fit, these authors indicated the range for the planet's orbital period to be [4,6)\,AU, therefore setting a lower limit at 4\,AU but no clear upper limit, and guessing a the most likely value to lie around 5.5\,AU.

\par
In order to constrain definitively the orbital solutions of this complex system, a long-term follow-up was required. The many free parameters called also for a high measurements density. Last but not least, small signatures, like that of the innermost planet, required high instrumental precision and the effect of stellar pulsations to remain below the 1\,m\,s$^{-1}$. We decided therefore to continue a precise radial velocity follow-up of this star with HARPS and adopted a strategy to minimize the pulsation signal. Finally, we have combined the obtained HARPS data with CORALIE and UCLES data to obtain a time coverage as complete as possible.\footnote{The stellar characteristics of $\mu$\,Ara are not described here, since it is not within the scope of this paper to discuss the host star. We refer instead to already published data in \citet{Butler:2001} and  \citet{Santos:2004-b}.}

%
\section{Description of the data set}                                

More than 25\% of the known extrasolar planets populate systems with at least two planets.  In most of these systems, the various planets were discovered sequentially, starting with the planet inducing the highest radial-velocity amplitude at short or intermediate period. The following planets unveiled themselves when, after a while, the residuals became high compared to the expected measurement precision and showed some structure as a function of time.

\par
$\mu$\,Ara clearly falls under this category of systems. Both increased data amount and the enlarged time span lead to the discovery of two additional companions. While the orbit of $\mu$\,Ara\,c is well constrained by the HARPS data, some uncertainties remain on the parameters of the third companion. With increasing number of planets, and thus free fit parameters, the amount of possible solutions increases drastically, and the only way to constrain the orbits is to acquire many new and precise data points.

\subsection{High-precision radial-velocity follow up of $\mu$\,Ara with HARPS}       

We have pursued this objective by observing  $\mu$\,Ara regularly with HARPS. In the time span between May 2004 (asteroseismology run) and July 2006, we have measured $\mu$\,Ara  on 85 different nights. One additional measurement had been recorded in summer 2003, before HARPS entered in operation. Apart from this older measurement, each data point consists of a small series of measurements covering a total integration time of 12 minutes. A typical observation scheme foresees 4 exposures of 3 minutes. The SNR reached in one single frame is of the order of 250 per extracted pixel, corresponding to a photon-noise limited accuracy of about 0.25\,m\,s$^{-1}$. Therefore, the photon-noise contribution of a complete observation to the total radial-velocity error lies between 0.15\,m\,s$^{-1}$ and 0.20\,m\,s$^{-1}$. The radial-velocity errors given in the data Table\,\ref{ta:harrvdata} have however been assumed more conservatively: We added quadratically to the photon-noise error a constant term of 0.8\,m\,s$^{-1}$, which includes calibration errors, pulsation noise, stellar jitter, and other possible short and long-term effects. This error estimation is well confirmed by typical dispersion values obtained on stable stars (see \citet{Lovis:2005} and \citet{Pepe:2005}, although the recent discovery of three planets around HD\,69830 \citep{Lovis:2006} indicate that HARPS can provide long-term accuracy of the order of 0.3\,m\,s$^{-1}$ on very stable stars. The nightly averages of the HARPS radial-velocity data and the estimated errors are shown in Table\,\ref{ta:harrvdata}. Note that for the asteroseismology night we have adopted an error value of 0.5\,m\,s$^{-1}$. This choice is motivated by the high number of data points which averages out completely any possible oscillation, guiding and atmospheric noise. This value is also motivated by the fact that when fitting HARPS data alone recorded in the period from June to August 2004 \citet{Santos:2004-b} obtain a dispersion of about 0.45\,m\,s$^{-1}$ on the asteroseismology-night data .

\subsection{Long-term CORALIE data}       

The HARPS data are of excellent quality but cover only the last two observation years. Consequently, they would not constrain possible planets with longer orbital periods. We have therefore used our CORALIE data which provide measurements as old as 8 years. The 39 CORALIE measurements acquired in 15-minutes exposures show typical photon error of 1-2\,m\,s$^{-1}$, to which we have added  quadratically a calibration noise of 3\,m\,s$^{-1}$. In contrast to HARPS, stellar pulsation and jitters does not significantly add to the error and can be neglected. It should be noted here that the CORALIE data have been reduced with the HARPS data-reduction software which has been recently adapted for CORALIE. However, some of the very first calibration frames did not pass the severe quality control, mainly because of the calibration exposures not yet being optimized in the first months of CORALIE operation. We have therefore chosen to remove these measurements from our data set. These data are not presented here and have not been used for the orbital fit.

Although the CORALIE data points are of lower quality compared with HARPS, new measurements have been carried out recently, in order to ensure good overlap with the HARPS data -- and thus a good RV-offset determination. The CORALIE radial-velocity data and the estimated errors are shown in Table\,\ref{ta:corrvdata}.

\subsection{UCLES data}       
As mentioned above, the total amount of data points is critical for constraining unambiguously the orbital parameters of a multi-planetary system. In order to complete our data set in terms of time coverage and sampling, we have integrated in this study the UCLES data on  $\mu$\,Ara published by \citet{McCarthy:2004}, to which we refer for more details.

%
\section{The $\mu$\,Ara multi-planetary system}                              

\subsection{Combining $\mu$\,Ara radial-velocity data}

The complete data set covers about 8 years of observations and contains 171 nightly-averaged measurements of quite different precision depending on the respective instrument. The various instruments cover different periods and show week overlap in time. One difficulty consists in assigning the correct error bars to the measurements, in particular of one instrument relative to the other. In fact, wrong error estimation, and thus weights, could lead to emphasizing one time domain more than another. Even worse, an instrument which covered a short time span with high density will probe short-period planets, while it would be more sensitive to long-period planets if its observation had covered a longer time span.
\par
The computation of the HARPS and CORALIE measurement errors has been performed as described in the previous section. For the UCLES data we have used the error values indicated in \citet{McCarthy:2004}. Effects due to jitter or the p-mode pulsation of $\mu$\,Ara, as well as long-term instrumental accuracy cannot be taken into account precisely, especially at the level of 1\,m\,s$^{-1}$ precision. We consider therefore the used error bars for the various instruments as realistic estimates although not fully reliable. 
\par
The mean radial velocity provided by the three instrument is expected to be quite different between the various instrument, especially between the UCLES instrument on one hand and CORALIE and HARPS on the other. The reason is that the UCLES data have been recorded using the iodine-cell technique, which does not provide the stellar velocity offset, since it measures radial-velocity changes compared to some reference spectrum. Instead, the thorium technique used on HARPS and CORALIE measures the stellar radial velocity relative to the laboratory rest frame. Although these two latter instruments are operated in the same mode and use the same pipeline, we can not exclude a priori some small instrumental offset due, for instance, to the different optical configuration, resolution, cross-dispersion etc. We have therefore decided to leave the offsets of the data of each instrument as additional free parameters. The two additional degrees of freedom are fully acceptable, if we consider the large number of data points.

\subsection{Search for the orbital solution}
\label{se:solution}
Despite the large amount of measurements, the search for the best orbital solution remains challenging: The irregular time sampling and the time span, in combination with the possibly large number of free fit parameters, produce a complex and rapidly varying $\chi^2$ function. The major risk is to remain trapped in a local minimum which does not correspond to the solution with absolute minimum $\chi^2$. On the other hand, we might come up with different types of solutions showing insignificantly different $\chi^2$. 

\par
In order to explore the entire parameter range and to find the best solution in terms of residuals minimization, we have used the \emph{Stakanof} software package recently developed at the Geneva Observatory \citep{Tamuz:2006}. \emph{Stakanof} combines genetic algorithms for exploring efficiently the entire parameter space with algorithms allowing for fast convergence into local minima. \emph{Stakanof} delivers a full set of solutions (same $\chi^2$ within the uncertainty range defined by the measurement errors) in less than 20 minutes on an ordinary personal computer. \emph{Stakanof} performs however only fitting of sum of any number of Keplerians but does not take into account planet-planet interaction in its model.

\subsubsection{From a two-planet to a four-planet system}
\begin{figure}
\includegraphics[width=\columnwidth]{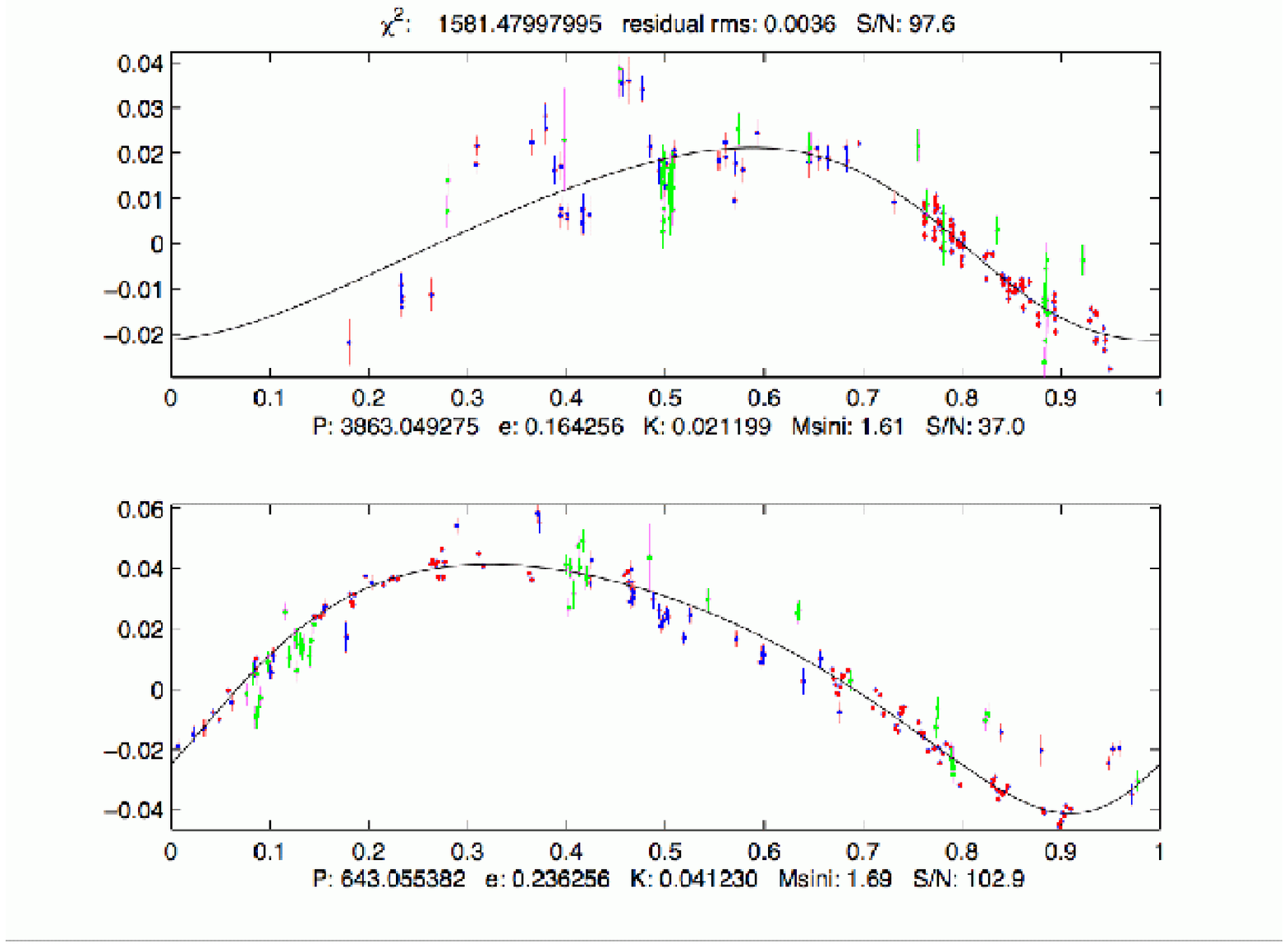}
\centering
\caption{
Phase-folded radial-velocity measurements and best fit for the two planets in the two-planet solution. For every planet the contribution by the other planets has been subtracted. On the upper panel showing the long-period planet a time-correlated structure becomes evident. HARPS data are plotted in red, UCLES data in blue, and CORALIE data in green.}
\label{fi:two_planets}
\end{figure}

In the present case we have started our exploration by assuming only 2 planets in the system: The well known 1.3\,M$_{Jup}$ planet orbiting $\mu$\,Ara in about 640\,days and a planet with a long period, as motivated by the observed trends. For simplicity, and in order to disentangle various questions, we did not fit for the short-period 14\,M$_{\oplus}$ planet on the 9.6\,days orbit. The orbit-search program \emph{Stakanof} immediately finds the well-known planet on an eccentric orbit ($e=0.24$) of 643\,days. A second planet is found on an orbit of about 3863\,days and shows eccentricity of $e=0.16$. As we will see later, the $\chi^2=1581$ and the weighted dispersion of the residuals of 3.6\,m\,s$^{-1}$ are quite high. In particular, the residuals to the data show time-correlated structure, as can been seen from the upper panel of Figure\,\ref{fi:two_planets}. We think that this solution is not satisfactory, and that a fourth planets, in addition to the short-period low-mass planet, must be present in the system.

Therefore, we have explored a solution which foresees the presence of four planets. Besides the 640\,days-period and the short-period planets, and a planet with a period longer than 2000\,days, we guessed the presence of an additional planet with intermediate period of about 300\,days. Despite any possible a priori knowledge we might have introduced, we decided to leave all orbital parameters for all the planets completely free, and let \emph{Stakanof} search for a solution with four planets in the full parameter space.

\par
In contrast to our expectations, the program converges rapidly towards a class of very good solutions with weighted dispersion values around 1.75\,m\,s$^{-1}$ and $\chi^2=442$. The system contains the known planet $\mu$\,Ara\,b at $P=643.3$\,days, with $e=0.128$,  $K=37.8$\,m\,s$^{-1}$ and $M=1.68$\,M$_{Jup}$, as well as the Neptune-mass planet ($\mu$\,Ara\,c, with 10.5\,M$_{\oplus}$ in this fit) at 9.64\,days, with $e=0.171$ and $K=3.1$\,m\,s$^{-1}$. In addition, the fits confirms our guess by finding a new 0.522\,M$_{Jup}$ planet at 310.6\,days. The eccentricity of this planet is low  $e=0.067$ and the amplitude is $K=14.9$\,m\,s$^{-1}$. Finally, also a clear signal at longer period is found. The best fit delivers a planet with M$=1.81$\,M$_{Jup}$, P$=4205.8$\,days, $e=0.098$ and $K=21.8$\,m\,s$^{-1}$.

\subsubsection{Exploring the parameter space}
The \emph{Stakanof} software finds also other solutions with a $\chi^2$ only slightly higher (within 10\%), but with parameters which are quite different for the outer planet. In fact the orbital period can easily vary from 3200 to 5000\,days, and its mass correspondingly. While the orbital parameters of the three inner planets  remain essentially constant, the orbit of the outermost planet can vary considerably, mainly because the data points have not yet covered a full orbital cycle. We hence conclude that although the existence of the outermost planet is out of doubt, its period and mass are fairly uncertain. Furthermore, the orbital parameters of this planets slightly influence the eccentricity and the mass of the third planet, which may in turn change the dynamic of the system. An investigation of the system stability, as done in the next section, may therefore contribute further inputs to the characterization of the system, or better, help us excluding unstable systems from our solution space. 

\par
Interestingly, still using \emph{Stakanof} we have also found a completely different solution with a $\chi^2$ identical to the previously presented best fit. This system also has four planets with the inner planet at 9.64 days and the outer one at about 2741\,days, but, in contrast to the previous solution, the intermediate pair of planets are trapped in a 1:1 mean motion resonance around 600 days; a pair of Trojan planets. As shown in the next section, we can exclude however this solution from the set of the possible ones, because it reveals to correspond to an highly unstable configuration. 

\subsubsection{Characteristics of the  $\mu$\,Ara four-planet system}
\begin{figure}
\includegraphics[clip=true, width=\columnwidth]{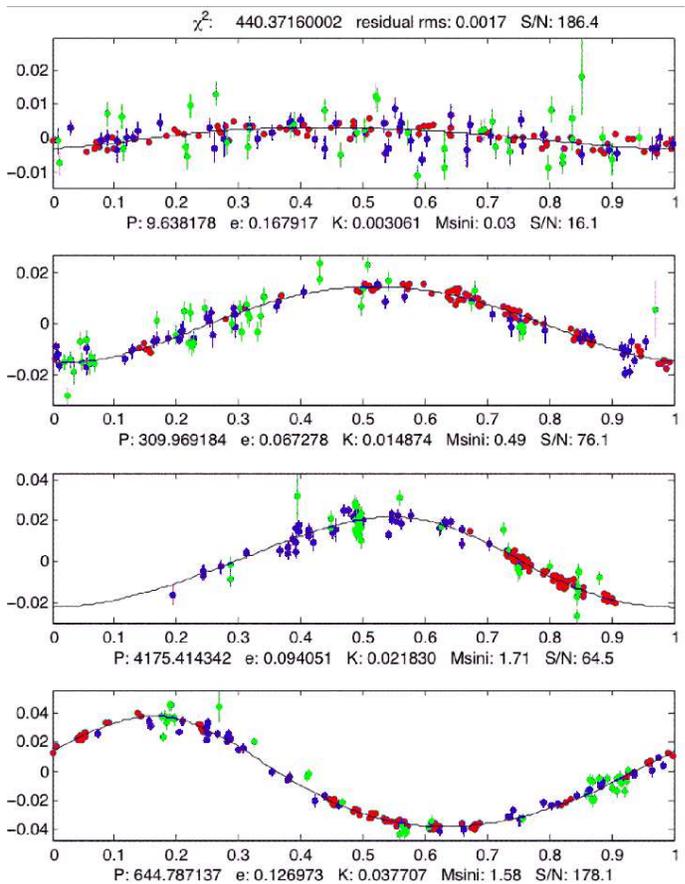}
\centering
\caption{
Phase-folded radial-velocity measurements and best fit for any of the four planets. For every planet the contribution by the remaining three planets has been subtracted. HARPS data are plotted in red, UCLES data in blue, and CORALIE data in green. Note that on the third panel showing the longest-period planet the HARPS data clearly indicate a flattening of the radial velocity curve. The recent HARPS data consequently constrain the orbital period of this planet.}
\label{fi:allinphase}
\end{figure}

\begin{figure}
\includegraphics[width=\columnwidth]{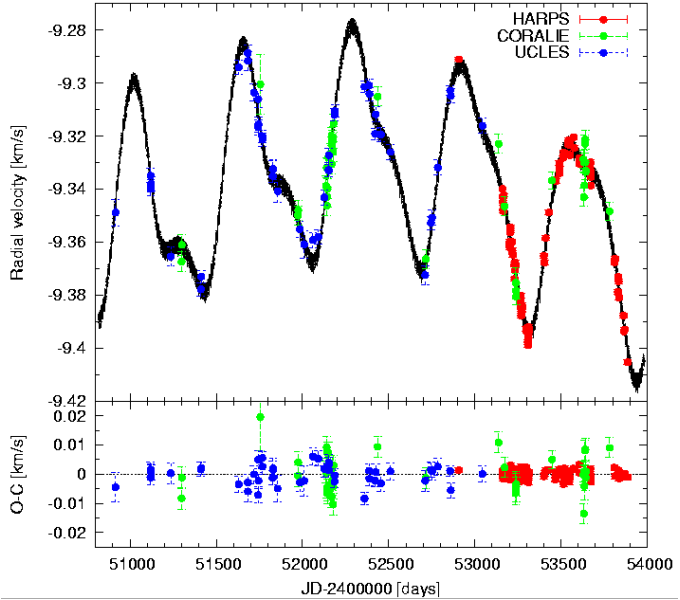}
\centering
\caption{
Global solution containing 4 planets and covering the full time range of observations. The lower plot shows the residuals of the fitted function to the actual measurements. The fit takes into account planet-planet interaction. The weighted dispersion of the residuals is 1.75\,m\,s$^{-1}$. }
\label{fi:global_solution}
\end{figure}

\begin{figure}
\includegraphics[width=\columnwidth]{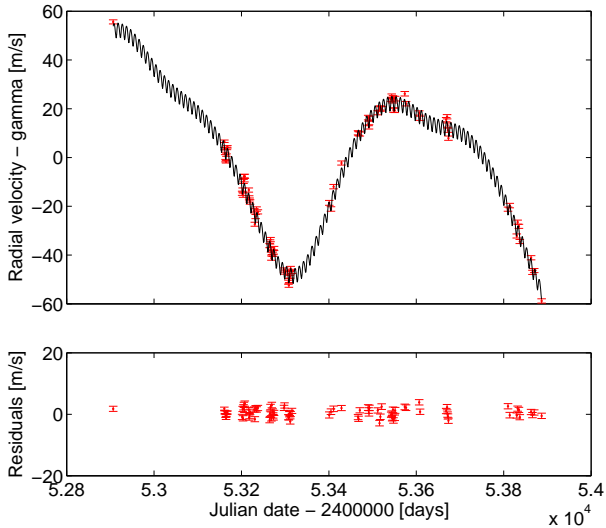}
\centering
\caption{
The figure shows a zoom on the time span covered by the HARPS data. It must be note that HARPS data constrain in an exceptional way the three shorter-period orbit. It particular, these complex curve would be incompatible even with a two-planet solution.}
\label{fi:harps_intime}
\end{figure}

In Figure\,\ref{fi:allinphase} we present our best 4-planets Keplerian fit. The plot shows the phase-folded data and the best fit for the four planets individually. In each of the plots the contribution by the remaining three planets has been subtracted. For the detailed description of the orbital parameters of the planets, we used a n-body fit that takes into account the planet-planet interaction (Table\,\ref{ta:orb}). Since the observational data cover only a short period of time (about 8.5 years), the orbital elements obtained with a n-body fit do not present significant differences with respect to the one obtained with \emph{Stakanof}. Figure\,\ref{fi:global_solution} shows the global solution, i.e. the combination of the motion of the four planets, as a function of time, and over-plotted with the actual measurements.  Figure\,\ref{fi:harps_intime} represents a zoom on the time span covered by HARPS data. Only the HARPS data are shown in this figure to emphasize how well the curve is constrained by these observations. Even by covering less than two cycles of the $P=640$\,days planet, it becomes clear that two other planets (in addition to the very-short period planet) must be present. In particular, the flattening of the curve after the first relative maximum is a clear indication for the presence of the previously unknown companion with $P=310$\,days.

\begin{figure}
\includegraphics[width=\columnwidth]{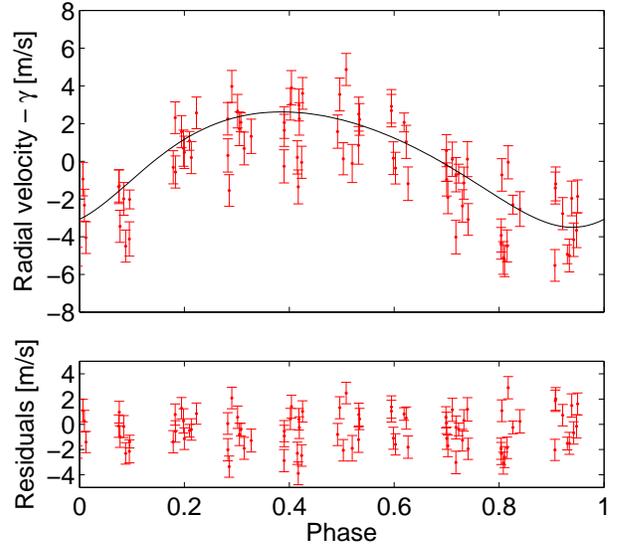}
\centering
\caption{
The upper panel shows the phase-folded HARPS data and the best Keplerian best fit to the $P=9.64$\,days-planet after subtracting the contribution of the other planets. The lower panels shows the residuals of the data to the best fit. The dispersion of the HARPS residuals is 1.41\,m\,s$^{-1}$  \emph{rms}.  }
\label{fi:harps_inphase}
\end{figure}

The Neptune-mass planet $\mu$\,Ara\,c is hardly recognized on the UCLES and CORALIE data. The semi-amplitude of 3.1\,m\,s$^{-1}$ is only constrained by the HARPS measurements. We have therefore plotted in Figure\,\ref{fi:harps_inphase} the phase-folded HARPS data only, after subtracting the contribution of the three other planets, in order to make the low-mass planet amplitude variation to appear clearly. the scatter of the residuals is 1.41\,m\,s$^{-1}$  \emph{rms}, which must be compared to UCLES and CORALIE data, with their 3.34\,m\,s$^{-1}$  \emph{rms} and 6.09\,m\,s$^{-1}$  \emph{rms}, respectively (Table\,\ref{ta:ins}). It is important to note that the HARPS data cover a a time span of 530\,days, even without accounting for the very first measurements recorded one year before the asteroseismology series. This demonstrates that the oscillations remained fully in phase during the covered period, and that it is therefore very unlikely that they are due to spots and activity of the star, as also the absence of correlation between radial velocities and line bisector proves. It is worth noticing that the amplitude of $\mu$\,Ara\,c is smaller than that reported in the discovery paper of $K=4.1$\,m\,s$^{-1}$. This difference cannot be explained by the error bar only. The main difference come from the fact that at the time of its discovery the longest-period planet was poorly characterized and the $P=310.6$\,days planet was even not known, both leading to a wrong estimation of the local "slope" in the radial-velocity data.

We estimated that planet-planet interaction cannot be neglected in the present system, since the second and the third planets are rather close to each other. We have therefore re-fitted the radial-velocity data by including interaction in our model and starting from the values computed from the Keplerian fit. The obtained orbital elements (Table\,\ref{ta:orb}) are very similar to the Keplerian solution. Since they are more general, we shall adopt these values and use them in the following as baseline for our study of the system dynamics.

\begin{table}
\centering
\caption{Orbital parameters and inferred planetary characteristics derived from
the best-fit orbital solution with planetary perturbations for the four planets
around $\mu$\,Ara at JD = $2\,453\,000 $ and keeping the inclination fixed at
$90^\circ $. For this fit we have got a weighted
dispersion of 1.75\,m\,s$^{-1}$ and $ \chi^2 = 442 $. The reduced  $ \chi^2_{red}$ of this best fit is 1.729.}  
\label{ta:orb}

\begin{tabular}{l@{\hspace{0.2cm}}lr@{\hspace{0.2cm}$\pm$\hspace{0.2cm}}l}
\hline
  \multicolumn{2}{l}{\bf Parameter} 
& \multicolumn{2}{c}{\bf $\mu$\,Ara\,b} \\
\hline
$P$ &$[$days$]$ & 643.25 & 0.90\\
$T$ &[JD] & 2452365.6 & 12.6 \\
$\lambda_0$ & [deg] & 17.6 & 0.4 \\
$e$ & & 0.128 & 0.017 \\
$\omega$ &[deg] & 22.0  & 7.0\\
$K$  &[m\,s$^{-1}$] & 37.78 & 0.40\\
\hline
$a_1\sin i$ & [AU] & \multicolumn{2}{c}{2.215$\cdot 10^{-3}$}\\
$f(m)$ &$\mathrm{[M_{\odot}]}$ 
  & \multicolumn{2}{c}{3.504$\cdot 10^{-9}$} \\
$m_{1}$ &$\mathrm{[M_{\odot}]}$ 
        & \multicolumn{2}{c}{1.08} \\
$m_{2}\,\sin i$ &$\mathrm{[M_{\rm Jup}]}$ 
        & \multicolumn{2}{c}{1.676} \\
$a$ &[AU] & \multicolumn{2}{c}{1.497}\\
\hline
\end{tabular}

\vspace{0.8 cm}

\begin{tabular}{l@{\hspace{0.2cm}}lr@{\hspace{0.2cm}$\pm$\hspace{0.2cm}}l}
\hline
  \multicolumn{2}{l}{\bf Parameter} 
& \multicolumn{2}{c}{\bf $\mu$\,Ara\,c} \\
\hline
$P$ &$[$days$]$ & 9.6386 & 0.0015\\
$T$ &[JD] & 2452991.1 & 0.4 \\
$\lambda_0$ & [deg] & 183.4 & 2.9 \\
$e$ & & 0.172 & 0.040\\
$\omega$ &[deg] & 212.7  & 13.3 \\
$K$  &[m\,s$^{-1}$] & 3.06 & 0.13\\
\hline
$a_1\sin i$ & [AU] & \multicolumn{2}{c}{2.670$\cdot 10^{-6}$}\\
$f(m)$ &$\mathrm{[M_{\odot}]}$ 
  & \multicolumn{2}{c}{2.734$\cdot 10^{-14}$} \\
$m_{1}$ &$\mathrm{[M_{\odot}]}$ 
        & \multicolumn{2}{c}{1.08} \\
$m_{2}\,\sin i$ &$\mathrm{[M_{\rm Jup}]}$ 
        & \multicolumn{2}{c}{0.03321} \\
$a$ &[AU] & \multicolumn{2}{c}{0.09094}\\
\hline
\end{tabular}

\vspace{0.8 cm}

\begin{tabular}{l@{\hspace{0.2cm}}lr@{\hspace{0.2cm}$\pm$\hspace{0.2cm}}l}
\hline
  \multicolumn{2}{l}{\bf Parameter} 
& \multicolumn{2}{c}{\bf $\mu$\,Ara\,d} \\
\hline
$P$ &$[$days$]$ & 310.55 & 0.83 \\
$T$ &[JD] & 2452708.7 & 8.3 \\
$\lambda_0$ & [deg] & 167.3 & 1.7 \\
$e$ & & 0.0666 & 0.0122 \\
$\omega$ &[deg] & 189.6 & 9.4 \\
$K$  &[m\,s$^{-1}$] & 14.91 & 0.59 \\
\hline
$a_1\sin i$ & [AU] & \multicolumn{2}{c}{4.247$\cdot 10^{-4}$}\\
$f(m)$ &$\mathrm{[M_{\odot}]}$ 
  & \multicolumn{2}{c}{1.060$\cdot 10^{-10}$} \\
$m_{1}$ &$\mathrm{[M_{\odot}]}$ 
        & \multicolumn{2}{c}{1.08} \\
$m_{2}\,\sin i$ &$\mathrm{[M_{\rm Jup}]}$ 
        & \multicolumn{2}{c}{0.5219} \\
$a$ &[AU] & \multicolumn{2}{c}{0.9210}\\
\hline
\end{tabular}

\vspace{0.8 cm}

\begin{tabular}{l@{\hspace{0.2cm}}lr@{\hspace{0.2cm}$\pm$\hspace{0.2cm}}l}
\hline
  \multicolumn{2}{l}{\bf Parameter} 
& \multicolumn{2}{c}{\bf $\mu$\,Ara\,e} \\
\hline
$P$ &$[$days$]$ & 4205.8 & 758.9 \\
$T$ &[JD] & 2452955.2 & 521.8 \\
$\lambda_0$ & [deg] & 61.4 & 9.2 \\
$e$ & & 0.0985 & 0.0627 \\
$\omega$ &[deg] & 57.6 & 43.7 \\
$K$  &[m\,s$^{-1}$] & 21.79 & 2.30\\
\hline
$a_1\sin i$ & [AU] & \multicolumn{2}{c}{8.382$\cdot 10^{-3}$}\\
$f(m)$ &$\mathrm{[M_{\odot}]}$ 
  & \multicolumn{2}{c}{4.441$\cdot 10^{-9}$} \\
$m_{1}$ &$\mathrm{[M_{\odot}]}$ 
        & \multicolumn{2}{c}{1.08} \\
$m_{2}\,\sin i$ &$\mathrm{[M_{\rm Jup}]}$ 
        & \multicolumn{2}{c}{1.814} \\
$a$ &[AU] & \multicolumn{2}{c}{5.235}\\
\hline
\end{tabular}

\end{table}

\begin{table}
\centering
\caption{Table of the instrument-specific characteristics}
\label{ta:ins}
\begin{tabular}{l@{\hspace{0.cm}}lr@{\hspace{0.3cm}\hspace{0.cm}}lr@{\hspace{0.cm}$\pm$\hspace{0.cm}}lr@{\hspace{0.cm}$\pm$\hspace{0.cm}}l}
\hline
  \multicolumn{2}{l}{\bf } 
& \multicolumn{2}{c}{\bf HARPS} & \multicolumn{2}{c}{\bf UCLES} & \multicolumn{2}{c}{\bf CORALIE} \\
\hline
$N_{\rm meas}$ & & \multicolumn{2}{c}{86} & \multicolumn{2}{c}{45} & \multicolumn{2}{c}{39}\\
$\delta\gamma$  &[km\,s$^{-1}$] & 0.0 & ref. & 9.3306 & 0.0008 & 0.03366 & 0.0010\\
$\sigma (O$-$C)$\hspace{.25cm}  & [m\,s$^{-1}$] & \multicolumn{2}{c}{1.41} & \multicolumn{2}{c}{3.34} & \multicolumn{2}{c}{6.01}\\
\hline
\end{tabular}
\end{table}

%
\section{Stability of the system}

\begin{figure}
\includegraphics[width=\columnwidth]{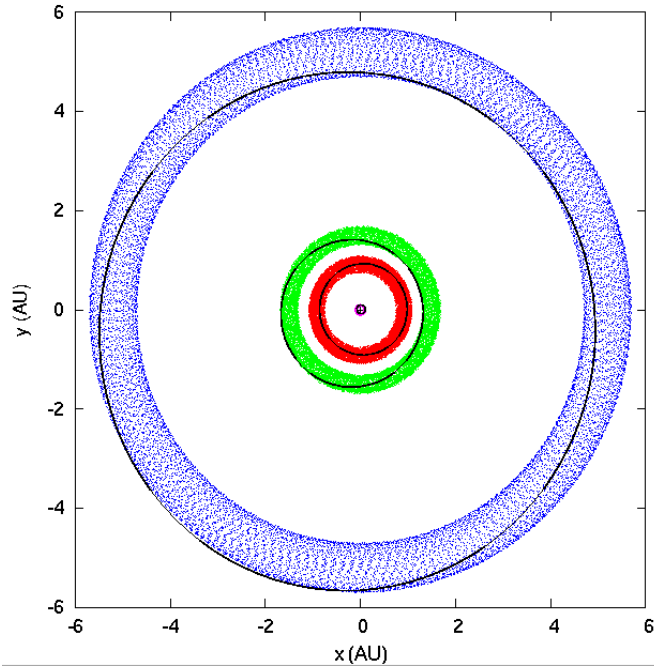}
\centering
\caption{
Orbital evolution of the four planets over one million years starting with the orbital
solution from Table\,\ref{ta:orb} and an inclination of $90^\circ$. The panel
shows a face-on view of the system; x and y are spatial coordinates in a frame
centered on the star. Present orbital solutions are traced with solid lines and
each dot corresponds to the position of the planet every 50 years. The
semi-major axes are constant, and the eccentricities undergo small variations
($0.09 < e_b < 0.13 $; $ 0.16 < e_c < 0.21 $; $ 0 <
e_d < 0.19 $;  $ 0.08 < e_{(e)} < 0.11 $).}
\label{fi:global_orbits}
\end{figure}

Running a simulation over one million years using the initial conditions from Table\,\ref{ta:orb}, we observe that the orbits of the two intermediate-period planets, $\mu$\,Ara\,b and $\mu$\,Ara\,d, can come together as close as 0.62~AU (Figure\,\ref{fi:global_orbits}). Consequently, the inner region of this system is subject to important gravitational interactions that may destabilize the orbits. Indeed, tracking the dynamical evolution for longer periods of time, the system is destroyed after 76~million years. 

The estimated age of the $\mu$\,Ara system is about 6.4\,Gyr \citep{Saffe:2005},
indicating that the previous orbits were not completely well determined. One reason is that the fitted parameters still present some uncertainties around the best fitted value. This is particularly true for the
outermost planet, whose orbit is not yet completely constrained because the orbital
period is longer than the present 8.5 years of observational data.
Moreover, there may exist additional planets in the system that are disturbing
the present solution, but that could not be detected yet.
We should thus consider that the set of parameters given in Table\,\ref{ta:orb}
constitutes the best determination one can do so far in terms of minimum $\chi^2$, and we will therefore search for more stable solutions in its vicinity.

\begin{figure}
\includegraphics[width=\columnwidth]{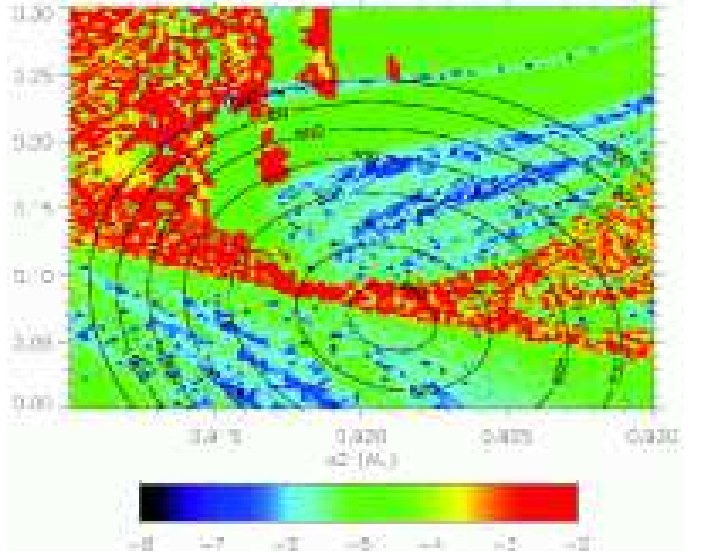}
\centering
\caption{Global view of the dynamics of the $\mu$\,Ara
    system for variations of the 
    eccentricity with the semi-major axis of $\mu$\,Ara\,d. 
    The color scale is the stability index ($D$) obtained through a frequency 
    analysis of the longitude of the outer planet over two 
    consecutive time intervals of 1000 yr.
    Red areas correspond to high orbital diffusion
    (instability) and the blue ones to low diffusion (stable orbits).
    Labeled lines give the value of $ \chi^2 $ obtained for
    each choice of parameters. There are two different types of blue zones where the system can be
    stabilized, one of them corresponding to the 2:1 mean
    motion resonance (for $ a_d > 0.915 $~AU and $ e_d > 0.15 $).}
\label{fi:stab1}
\end{figure}

Since the orbits of the innermost planet $\mu$\,Ara\,c and the third planet
$\mu$\,Ara\,b are well established, with small standard errors,
we have kept the parameters of these two planets constant.
We also did not change the inclination of the orbital planes, keeping
both at $90^\circ$.  For the second planet, $\mu$\,Ara\,d, we let $a$,
$\lambda$, $e$ and $\omega$ vary. 
Typically, as shown in Figure\,\,\ref{fi:stab1}, we have fixed $\lambda$ and $\omega$ to
specific values, and have spanned the $(a,e)$ plane of initial
conditions with a step size of  $0.005$~AU for $a$ and $0.05$ for $e$, in
a similar way as it was done for the {\small HD}\,202206 system \citep{Correia:2005}.
For each initial condition, the orbits of the three outer planets are integrated over 2000 years with the symplectic integrator SABA4 of \citet{Laskar:2001}, using a step size of $0.02$ year. The innermost planet has a very small influence on the stability of the rest of the system, removing it allows us to use a step size as big as $0.02$ year. The stability of the orbit is then measured by 
frequency analysis \citep{Laskar:1990,Laskar:1993}. Practically, a refined 
determination of the mean motion $n_d, n_d'$ of the second planet 
is obtained over two consecutive time intervals of $T=1000$ years, 
and the measure of  the difference $D = | n_d-n_d'|/T$ 
(in deg/yr${}^2$ in Figure\,\,\ref{fi:stab1}) is a measure of the chaotic diffusion
of the trajectory. It should be close to zero for a regular solution and 
high values will correspond to strong chaotic motion.
In the present case a regular motion will require  $ D < 10^{-6}$.
Labeled lines of Figure~\ref{fi:stab1} give the value
of $\chi^2$ obtained for each choice of parameters.

We find that the vicinity of $\mu$\,Ara\,d ($P=310.6$\,days) is somewhat chaotic (red
regions on Figure\,\,\ref{fi:stab1}) and almost no regular motion is found. Because of the proximity of $\mu$\,Ara\,b, the chaotic
behavior was expected. Studying the $(a,e)$ plane of initial conditions (Figure\,\,\ref{fi:stab1}), we find three main regions with qualitatively different dynamical behavior. A first region is delimited by 
\begin{equation} 
e_d < e_d^{(lim)} \approx 0.2 \times \left( \frac{0.937-a_d}{0.937-0.89} \right) \ .
\end{equation}
The fit from Table\,\ref{ta:orb} lies at the edge of this region, as shown by the $\chi^2$ level curves. Orbits in this particular region are chaotic, and the second and third planets experience close encounters within a few tens of million years. More regular orbits can be found in the small blue patch with $a_d$ roughly between $0.913$ and $0.92$, and $e_d < 0.05$, where the system can be stabilized. Above this region, and for $a_d<0.915$~AU, orbits are highly unstable, most of them lost in less than $2000$ years (white dots). Finally, for $a_d > 0.915$~AU, we find the region corresponding to the 2:1 mean motion resonance island (in Figure\,\,\ref{fi:stab1} we actually see only half of the island, as it expands up to $a_d\approx0.975$~AU). 
The main resonant argument is given by:
\begin{equation}
\theta = \lambda_d - 2 \lambda_b + \omega_d \ ,
\end{equation}
with a libration period of about 30 years. One expects in such configuration (massive and eccentric planets close to each other) that the 2:1 mean motion resonance stabilizes the system if the two planets are also in an apsidal co-rotational state.
 

Beside the solution listed in Table\,\ref{ta:orb} we have found two more solutions
with an identical $ \chi^2 $ and residuals (Section\,\ref{se:solution}).
One of these solutions is very similar to the solution from Table\,\ref{ta:orb}
except for the outer planet (less constrained) for which we find $ a_{(e)} = 5.65
$~AU, $ e_{(e)} = 0.138 $, $ \lambda_{0{(e)}} = 55.85^\circ $, $ \omega_{(e)} = 39.71^\circ $ and $
K_{(e)} = 23.25 $~m/s, corresponding to a minimum mass of 2~$M_{Jup} $.
The dynamical behavior of this system is very similar to the one listed in
Table\,\ref{ta:orb}. Indeed, the modifications in the orbit of the outer planet do
not change the global picture of the inner planetary system. A numerical
long-term simultation is stable over about 14 million years (after which the
system is destroyed), but we can slightly modify the parameters of the $\mu$\,Ara\,d
in order to stabilize the system for longer time scales.

\begin{figure}
\includegraphics[width=\columnwidth]{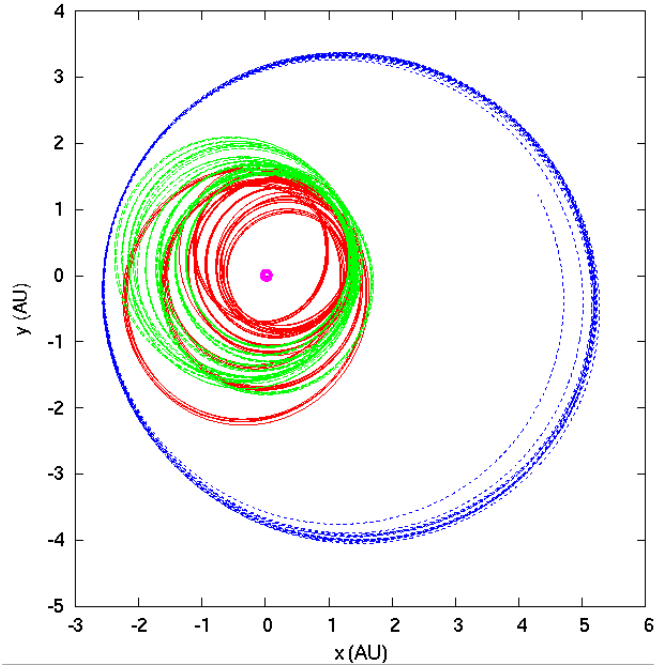}
\centering
\caption{
Orbital evolution of the four planets with a 1:1 resonance: inclination is equal
to $90^\circ$ and $ a_b = 1.47 $~AU, $ e_b = 0.128 $, $ \lambda_{0b} = 36.59^\circ $,
$ \omega_b = 255.68^\circ $, $ K_b = 62.00 $~m/s; $ a_c = 0.09 $~AU, $ e_c =
0.175 $, $ \lambda_{0c} = 176.60^\circ $, $ \omega_c = 206.04^\circ $, $ K_c = 3.20
$~m/s;  $ a_d = 1.42 $~AU, $ e_d = 0.122 $, $ \lambda_{0d} = 235.13^\circ $, $ \omega_d
= 358.80^\circ $, $ K_d = 27.94 $~m/s; and $ a_{(e)} = 3.93 $~AU, $ e_{(e)} =
0.350 $, $ \lambda_{0{(e)}} = 89.79^\circ $, $ \omega_{(e)} = 164.00^\circ $, $
K_{(e)} = 18.81 $~m/s. The panel shows a face-on view of the system; x and y are
spatial coordinates in a frame centred on the star. Due to the strong 
interactions, the orbits of the Trojan planets undergo large variations and the
system is destroyed in only 92 years. A dynamical analysis of the stability
around the Trojan planets also provides no alternative stable possibility.} 
\label{fi:stab2}
\end{figure}

The other solution provided by the genetic algorithm was a system of four
planets, where $\mu$\,Ara\,b and d were replaced by a pair of Trojan planets orbiting with the same
period as the former $\mu$\,Ara\,b in a 1:1 mean motion resonance. The innermost planet remained with the same orbital
parameters from the previous two solutions, but the outermost one is closer to
the inner system, orbiting just at $ 3.93 $~AU with $ e_{(e)} = 0.35$ and a minimum
mass of about 1.3~$M_{Jup} $ (see legend of Figure\,\ref{fi:stab2}). These parameters and the quality of the fit remain
essentially the same even when we conisder the planet-planet interactions.
However, due to the strong interactions, if we follow this system for a few
decades we see that the orbits of the Trojan planets undergo large variations
and the system is destroyed in less than a century due to a collision between
the two Trojan planets.
(Figure\,\ref{fi:stab2}).
We also looked for stability in the vicinity of the Trojan planets, in a
similiar study as done for the other solutions (Figure\,\ref{fi:stab1}), but no
stable configurations were found when changing $a_d$, $e_d$ and $\omega_d$. 

\par
During our search with \emph{Stakanof}, we had encountered various other solutions similar to the best fit. The main difference between them was the period of the outer most planet, ranging from 3000 to 5000\,days. Given the large distance of this planet from the inner planets, its exact period will have no direct impact on the system stability. However, when fitting the radial-velocity data and minimizing $ \chi^2 $, different orbital elements for the outermost planet will also lead to slightly different orbital elements for the inner planets. This may in turn change the system dynamics, such that it would be necessary to investigate the parameters space vicinity for each of these solutions. We shall restrict ourselves in this paper to the statement that several stable solutions exists, which are compatible with the measured radial-velocity data. The presented solution (Table\,\ref{ta:orb}) belongs definitively to the group of the most probable ones (best fitted and more stable).

\par
In summary, we have found an orbital solution (Table\,\ref{ta:orb}). which is quite stable, but not completely locked over the time scales of the planetary system age. By slightly modifying the initial orbital elements one can find stable solutions in its vicinity, leading to only slightly increased $ \chi^2 $. There are two different zones where the system can be stabilized (Figure\,\ref{fi:stab1}), one of them corresponding to the 2:1 resonance. This result delivers to us the prove of existence for a solution which is compatible with the observed radial velocities on one hand, and with the system age on the other.

%
\section{About the naming of planetary companions}

Our results unambiguously demonstrate the presence of 3 planetary companions around $\mu$\,Ara: $\mu$\,Ara\,b, the long-known $P=643$\,days planet, $\mu$\,Ara\,c, the Neptune-mass planet on the short $P=9.64$\,days discovered during the HARPS asteroseismology run, and the new planet $\mu$\,Ara\,d,  in an orbit of $P=310.6$\,days. These planets have been fully characterized and only little uncertainty exists on their orbital parameters. We have therefore adopted the naming of these three planets corresponding to the time sequence of their discovery \emph{and} characterization.

\par
The orbital elements of of the outermost planet  asindicated by \citet{Jones:2002} first and revised by McCarthy later are not anymore consistent with the present (extended) data set. The recently acquired HARPS data have strongly contributed to constrain the system parameters. In contrast to the three well-characterized inner planets, the period remains however somewhat uncertain. Fitting the data yields still a wide range of possible solutions with very similar  $ \chi^2 $ values. The best-fit is obtained with a period of 4206\,days, which is slightly longer than the time span of the observations. Without degrading significantly the residuals, it is possible to find a solution within about 20\% of the period of the best solution.  Nevertheless, for the first this it is possible to exclude stellar mass, since the uncertainty on the mass given by the radial-velocity data is small. In addition, the data fit provides us with a dynamically stable solution. We have therefore chosen to name this planet  $\mu$\,Ara\,e in this paper.

\par 
We propose that a similar naming strategy is adopted for all new discoveries. It is not sufficient to indicate trends in the RV-data or loose orbital elements to assess the planetary nature of the companion. There is the risk also to provide wrong inputs to the theoreticians, who must have to rely on assessed results (within the measurements uncertainty) to produce models which can be compared with the observations. It is worth noting that the data we had acquired until January 2006 led us first to a 4-planet solution, which turned out to be dynamically unstable. Although the stability of the system is not sufficient to prove the correctness of the solution, it can be considered as a necessary condition to make it plausible. This led us to the decision to wait for additional HARPS data, in order to provide results with a higher confidence level.

%
\section{Conclusions}

In this paper we have characterized a system of planets around  $\mu$\,Ara to great detail. In addition to the already well-known planet  $\mu$\,Ara\,b on a 643\,days orbit and with a minimum mass of 1.68\,M$_{ Jup}$, we confirm the presence of the Neptune-mass at $P=9.64$\,days. The new data points constrain the mass of this planet to only 10.5\,M$_{\oplus}$. We have also presented the discovery and the characterization of a new, until date unknown planet on a $P=310.6$\,days orbit, $\mu$\,Ara\,d. This planet accounts alone for the previously anomalously high dispersion of the data points around the fitted orbit. The minimum mass of this planet is well constrained to 0.52\,M$_{ Jup}$. Finally we confirm the presence of a fourth companion, $\mu$\,Ara\,e, with a likely period $P=4206$\,days and probable mass of 1.81\,M$_{ Jup}$. The orbital period is however not yet fully constrained.
\par
It is worth mentioning that these results could be obtained in particular thanks to the superb quality of the HARPS data and to the great time coverage delivered by the UCLES and CORALIE data, which helped constraining the many free parameters of the orbital solution. The weighted \emph{rms} of 1.75\,m\,s$^{-1}$ obtained on 171 data points, over a time span of 8.5 years, and on a strongly-pulsating star, is exceptionally low and leaves for the moment only little space for a fifth companion.
\par
$\mu$\,Ara is the second system known to harbor 4 planetary companions. According to new theoretical models and the experience acquired from the latest discoveries, we can expect to find many more of these systems in future. Precise radial velocity machines like HARPS and the increasing time coverage of the various ongoing follow-up programmes may provide us with many new amazing results.

\begin{acknowledgements}
We are grateful to all technical and scientific collaborators of the HARPS Consortium, ESO Head Quarter and ESO La Silla who have contributed with their extra-ordinary passion and valuable work to the success of the HARPS project. We would like to thank the European RTN ``The Origin of Planetary Systems (PLANETS, contract number HPRN-CT-2002-00308) for supporting this project. This research has made use of the SIMBAD database, operated at CDS, Strasbourg, France. We finally acknowledge the support by Swiss National Science Foundation for its continuous support in our research work. 

\end{acknowledgements}

\bibliography{bibpepe}
\bibliographystyle{bibtex/aa}

\longtab{3}{ 
\begin{longtable}{l l l}
\caption[]{HARPS radial-velocity data of HD\,160691. The indicated radial-velocity values are weighted means of various exposures during a single night.} \label{ta:harrvdata}\\
\hline

\textbf{Julian date} & \textbf{RV} & \textbf{RV error} \\ 
\textbf{[T - 2'400'000]} & \textbf{[\,km\,s$^{-1}$]}&\textbf{[\,km\,s$^{-1}$]} \\ 
\hline\hline
\endfirsthead 

\caption{continued.}\\ 
\hline
\textbf{Julian date} & \textbf{RV} & \textbf{RV error} \\ 
\textbf{[T - 2'400'000]} & \textbf{[\,km\,s$^{-1}$]}&\textbf{[\,km\,s$^{-1}$]} \\ 
\hline\hline
\endhead 

\hline 
\multicolumn{3}{c}{\small\sl continued on next page}\\
\endfoot 

\hline 
\endlastfoot 

52906.51936	 &  -9.2909	 &  0.00084\\	
53160.72599	 &  -9.3398	 &  0.00050\\
53161.72780	 &  -9.3428 	 &  0.00050\\
53162.72601	 &  -9.3448 	 &  0.00050\\
53163.72588	 &  -9.3477 	 &  0.00050\\
53164.72576	 &  -9.3482 	 &  0.00050\\
53165.68275	 &  -9.3455 	 &  0.00050\\	
53166.78196	 &  -9.3427 	 &  0.00050\\	
53167.72693	 &  -9.3421 	 &  0.00050\\	
53201.61987	 &  -9.3611	 &  0.00085\\	 
53202.64137	 &  -9.3598	 &  0.00085\\	 
53203.61075	 &  -9.3619	 &  0.00085\\	
53204.63545	 &  -9.3563	 &  0.00085\\	
53205.56147	 &  -9.3542	 &  0.00085\\	
53206.63707	 &  -9.3556	 &  0.00084\\
53207.66322	 &  -9.3542	 &  0.00085\\	 
53216.79388	 &  -9.3598	 &  0.00085\\
53217.60908	 &  -9.3630	 &  0.00085\\
53218.56732	 &  -9.3639	 &  0.00084\\
53219.67984	 &  -9.3647	 &  0.00091\\
53229.49489	 &  -9.3683	 &  0.00084\\
53230.57363	 &  -9.3738	 &  0.00083\\
53232.49288	 &  -9.3707	 &  0.00086\\
53234.66632	 &  -9.3676	 &  0.00089\\
53237.49386	 &  -9.3688	 &  0.00086\\
53264.52049	 &  -9.3826	 &  0.00088\\
53265.50482	 &  -9.3811	 &  0.00088\\
53266.49340	 &  -9.3801	 &  0.00089\\
53267.50962	 &  -9.3844	 &  0.00088\\
53268.51030	 &  -9.3877	 &  0.00088\\
53269.51171	 &  -9.3853	 &  0.00090\\
53271.48362	 &  -9.3862	 &  0.00089\\
53272.48518	 &  -9.3855	 &  0.00088\\
53273.48732	 &  -9.3852	 &  0.00088\\
53274.52460	 &  -9.3838	 &  0.00084\\
53297.52223	 &  -9.3926	 &  0.00086\\
53298.50560	 &  -9.3935	 &  0.00085\\
53307.50815	 &  -9.3976	 &  0.00084\\
53308.49826	 &  -9.3989	 &  0.00084\\
53309.50581	 &  -9.3969	 &  0.00085\\
53310.52537	 &  -9.3941	 &  0.00085\\
53311.51144	 &  -9.3951	 &  0.00084\\
53312.52384	 &  -9.3920	 &  0.00085\\
53315.50462	 &  -9.3931	 &  0.00085\\
53400.88939	 &  -9.3651	 &  0.00100\\
53404.89478	 &  -9.3676	 &  0.00092\\
53410.89394	 &  -9.3584	 &  0.00090\\
53428.86141	 &  -9.3487	 &  0.00084\\
53466.78683	 &  -9.3363	 &  0.00087\\
53467.84365	 &  -9.3367	 &  0.00085\\
53470.82430	 &  -9.3368	 &  0.00085\\
53489.88910	 &  -9.3301	 &  0.00093\\
53490.85565	 &  -9.3325	 &  0.00092\\
53491.84209	 &  -9.3339	 &  0.00089\\
53492.82606	 &  -9.3304	 &  0.00089\\
53511.85243	 &  -9.3281	 &  0.00085\\
53515.78177	 &  -9.3265	 &  0.00090\\
53517.79817	 &  -9.3261	 &  0.00089\\
53520.80548	 &  -9.3265	 &  0.00092\\
53542.68983	 &  -9.3226	 &  0.00086\\
53543.70021	 &  -9.3213	 &  0.00089\\
53544.76392	 &  -9.3228	 &  0.00088\\
53545.77313	 &  -9.3231	 &  0.00101\\
53546.76437	 &  -9.3220	 &  0.00088\\
53547.67799	 &  -9.3270	 &  0.00089\\
53550.67900	 &  -9.3271	 &  0.00091\\
53551.67212	 &  -9.3223	 &  0.00096\\
53573.65657	 &  -9.3203	 &  0.00091\\
53576.61243	 &  -9.3244	 &  0.00091\\
53606.63942	 &  -9.3282	 &  0.00088\\
53608.57512	 &  -9.3304	 &  0.00087\\
53668.48502	 &  -9.3309	 &  0.00086\\
53670.54485	 &  -9.3297	 &  0.00087\\
53671.52852	 &  -9.3332	 &  0.00087\\
53672.53005	 &  -9.3353	 &  0.00086\\
53673.54838	 &  -9.3385	 &  0.00087\\
53809.91489	 &   -9.3676	 &  0.00087\\
53813.91743	 &   -9.3662	 &  0.00091\\
53829.88048	 &   -9.3787	 &  0.00084\\
53831.91068	 &   -9.3730	 &  0.00084\\
53834.90788	 &   -9.3747	 &  0.00084\\
53836.90174	 &   -9.3809	 &  0.00084\\
53862.80026	 &   -9.3876	 &  0.00086\\
53865.79772	 &   -9.3936	 &  0.00086\\
53869.81768	 &   -9.3928	 &  0.00085\\
53886.76366	 &   -9.4053	 &  0.00087\\

\end{longtable} 
}

\longtab{4}{ 
\begin{longtable}{l l l}
\caption[]{CORALIE radial-velocity data of HD\,160691. The errors are significantly higher than on HARPS data and are mainly dominated by photon noise.} \label{ta:corrvdata}\\
\hline

\textbf{Julian date} & \textbf{RV} & \textbf{RV error} \\ 
\textbf{[T - 2'400'000]} & \textbf{[\,km\,s$^{-1}$]}&\textbf{[\,km\,s$^{-1}$]} \\ 
\hline\hline
\endfirsthead 

\caption{continued.}\\ 
\hline
\textbf{Julian date} & \textbf{RV} & \textbf{RV error} \\ 
\textbf{[T - 2'400'000]} & \textbf{[\,km\,s$^{-1}$]}&\textbf{[\,km\,s$^{-1}$]} \\ 
\hline\hline
\endhead 

\hline 
\multicolumn{3}{c}{\small\sl continued on next page}\\
\endfoot 

\hline 
\endlastfoot 

51298.862816	 & -9.4013	 & 0.0036\\
51299.887396	 & -9.3949	 & 0.0038\\
51755.598346	 & -9.3344	 & 0.0113\\
51973.901056	 & -9.3838	 & 0.0036\\
51975.880104	 & -9.3819	 & 0.0037\\
52136.669644	 & -9.3727	 & 0.0037\\
52140.637318	 & -9.3658	 & 0.0038\\
52142.662767	 & -9.3802	 & 0.0037\\
52143.625355	 & -9.3660	 & 0.0037\\
52144.641898	 & -9.3768	 & 0.0037\\
52145.571119	 & -9.3738	 & 0.0036\\
52150.613056	 & -9.3620	 & 0.0036\\
52164.524007	 & -9.3603	 & 0.0036\\
52168.534181	 & -9.3541	 & 0.0033\\
52169.522324	 & -9.3646	 & 0.0037\\
52171.538214	 & -9.3557	 & 0.0039\\
52172.559328	 & -9.3573	 & 0.0035\\
52173.532866	 & -9.3560	 & 0.0033\\
52177.550376	 & -9.3597	 & 0.0037\\
52178.552546	 & -9.3544	 & 0.0042\\
52180.553288	 & -9.3491	 & 0.0033\\
52436.705549	 & -9.3390	 & 0.0035\\
52715.915800	 & -9.4003	 & 0.0036\\
53137.825826	 & -9.3568	 & 0.0035\\
53172.732248	 & -9.3804	 & 0.0035\\
53237.734580	 & -9.4093	 & 0.0057\\
53238.544750	 & -9.4093	 & 0.0036\\
53239.574868	 & -9.4124	 & 0.0031\\
53239.584890	 & -9.4144	 & 0.0031\\
53447.912310	 & -9.3706	 & 0.0031\\
53630.526063	 & -9.3627	 & 0.0032\\
53632.492165	 & -9.3770	 & 0.0033\\
53633.480736	 & -9.3638	 & 0.0034\\
53635.509067	 & -9.3725	 & 0.0046\\
53638.610152	 & -9.3568	 & 0.0034\\
53639.509958	 & -9.3639	 & 0.0034\\
53641.512632	 & -9.3552	 & 0.0036\\
53643.536254	 & -9.3674	 & 0.0042\\
53644.515441	 & -9.3670	 & 0.0033\\
53781.889598	 & -9.3823	 & 0.0033\\

\end{longtable} 
}

\end{document}